\documentclass{iaus}
\usepackage{graphicx}

\title[Dynamo in the Intra-Cluster Medium] 
{Dynamo in the Intra-Cluster Medium: Simulation of CGL-MHD Turbulent Dynamo}

\author[Santos-Lima \etal\ ]   
{R. Santos-Lima$^1$, 
E. M. de Gouveia Dal Pino$^1$,
A. Lazarian$^2$, 
\break G. Kowal$^1$, 
\and D. Falceta-Gon\c calves$^{3}$,
}

\affiliation{$^1$ IAG, Universidade de S\~ao Paulo, Rua do Mat\~ao 1226,  S\~ao Paulo 05508-090, Brazil
\\ email: {\tt rlima@astro.iag.usp.br} \\[\affilskip]
$^2$ Astronomy Department, University of Wisconsin, Madison, WI, USA
\\[\affilskip]
$^3$ NAC, Universidade Cruzeiro do Sul, Rua Galv\~ao Bueno
868,  S\~ao Paulo 01506-000, Brazil
}

\begin{document}

\maketitle

\begin{abstract}
The standard magnetohydrodynamic (MHD) description of the plasma in the hot, magnetized gas of the intra-cluster (ICM) medium is not adequate because it is weakly collisional. In such collisionless magnetized gas, the microscopic velocity distribution of the particles is not isotropic, giving rise to kinetic effects on the dynamical scales. These kinetic effects could be important in understanding the turbulence, as so as the amplification and maintenance of the magnetic fields in the ICM. It is possible to formulate fluid models for collisonless or weakly collisional gas by introducing modifications in the MHD equations. These models are often referred as kinetic MHD (KMHD). 
Using a KMHD model based on the CGL-closure, which allows the adiabatic evolution of the two components of the pressure tensor (the parallel and perpendicular components with respect to the local magnetic field), we performed 3D numerical simulations of forced turbulence in order to study the amplification of an initially weak seed magnetic field. We found that the growth rate of the magnetic energy is comparable to that of the ordinary MHD turbulent dynamo, but the magnetic energy saturates in a level smaller than of the MHD case. We also found that a necessary condition for the dynamo works is to impose limits to the anisotropy of the pressure.

\end{abstract}

\firstsection 
\section{Kinetic MHD description of a weakly collisional plasma}

The hypothesis underlying the MHD description of a plasma are not justified when the level of colisionality of the gas is negligible. Hence, a MHD model of the hot, magnetized gas of the ICM medium is not adequate because it is weakly collisional. There, the typical ion Larmor radius $\rho_i$ is much smaller that its mean free path $\lambda_i$. For instance, in the Hydra A cluster, $\rho_i \sim 10^5$ km and $\lambda_i \sim 10^{15}$ km (based on data presented in \cite[En{\ss}lin \& Vogt 2006]{EnsslinVoght2006}).

Nevertheless, it is still possible to formulate fluid models for collisonless or weakly collisional magnetized gas by introducing modifications in the MHD equations. The basic equations of these models are the ideal MHD equations for conservation of mass, momentum and the induction equation, with the pressure tensor having possibility of different components parallel and perpendicular to the magnetic field:
\begin{equation}
P_{ij} = p_{\perp}\delta_{ij} + (p_{\parallel} - p_{\perp}) b_{i} b_{j}
\end{equation}
where the $b_{i}$ are the components of the unitary vector parallel to the magnetic field. These models are often referred as kinetic MHD (KMHD). 

The lowest order closure (no heating conduction) to the set of macroscopic equations is given by (\cite[Kulsrud 1983]{Kulsrud1983}):
\begin{eqnarray}
 \frac{d}{dt} \left ( \frac{p_{\perp}}{\rho B}\right ) = 0, & \;\;\;\;\;\;\;\;\;\; & \frac{d}{dt} \left ( \frac{p_{\parallel} B^{2}}{\rho^{3}}\right ) = 0
\end{eqnarray}

The resulting model is called CGL-MHD approximation (\cite[Chew \etal\ 1956]{Chew_etal1956}). These equations of state declare for the conservation of the magnetic momentum of the particles and conservation of entropy of the gas. 

The kinetic MHD models reveal linear instabilities originating from the destabilization of the MHD-analogous waves, when the difference between the pressures components reaches some level. The growth rate of these instabilities, in the linear regime, are proportional to the wave number of the perturbation. 

We observe the spontaneous development of anisotropy in the pressure when we perform simulations of forced turbulence using the CGL-MHD model. The instabilities arising from these anisotropies give rise to small-scale structures, and therefore, a considerable amount of the kinetic and magnetic energy accumulates in the smallest scales of the simulation (limited by the numerical dissipation), changing the usual Kolmogorov power law of the power spectrum in the inertial range. In the absence of numerical dissipation, the CGL-MHD approximation would reveal its serious problem: the growth rate of the highest wavenumbers increases without limits.

\section{Anisotropy limits}

When the magnetic field changes in a frequency higher then the Larmor frequency of the ions, the assumption of conservation of angular momentum is not reasonable anymore, as pitch angle scattering should have place. Therefore, there is a limit to the maximum growth rate of the kinetic instabilities, above what the anisotropy of the pressure is reduced as a result of this pitch angle scattering. 

Based on kinetic considerations (see \cite[Sharma \etal\ 2006]{Sharma_etal2006} and references therein), we impose thresholds for the anisotropy of the pressure: 
\begin{eqnarray}
 \label{eqn:limitanis}
1 - \frac{p_{\perp}}{p_{\parallel}} - \frac{2}{\beta_{\parallel}} \: \lesssim \: \zeta, & \;\;\;\;\;\;\;\;\;\; & \frac{p_{\perp}}{p_{\parallel}} - 1 \: \lesssim \: \frac{2 \xi}{\beta_{\perp}}
\end{eqnarray}
where $\beta_{\parallel} = p_{\parallel} / (B^{2} / 8 \pi)$ e $\beta_{\perp} = p_{\perp} / (B^{2} / 8 \pi)$. Following \cite[Sharma \etal\ 2006]{Sharma_etal2006}, we use $\zeta = 0.5$ and $\xi = 3.5$.

When the anisotropy overcomes any of these thresholds, the pitch angle scattering acts to reduce the anisotropy back to the threshold. It imposes some limits to the growth rate of the kinetic instabilities.

\section{Numerical simulation of the CGL-MHD turbulent dynamo}

Using a Godunov-MHD code modified to evolve the CGL-MHD equations (see \cite[Kowal \etal\ 2010]{Kowal_etal2010}), we performed 3D numerical simulations of forced turbulence, starting with a weak seed magnetic field. The turbulence is forced in a periodic box by a random, non-helical, solenoidal force acting around a scale $2.5$ times smaller than one side of the box. In code units, the random velocity is kept close to unity, hence one turn-over time of the turbulence is $\approx 0.4$. The initial fields are uniform and have the following values: $\rho=1$ (density), $p_{\parallel}=p_{\perp}=1$, and $B=10^{-4}$ (magnetic field). The resolution employed is $64^3$.

\begin{figure}
\centering
\resizebox{9.0cm}{!}{\includegraphics{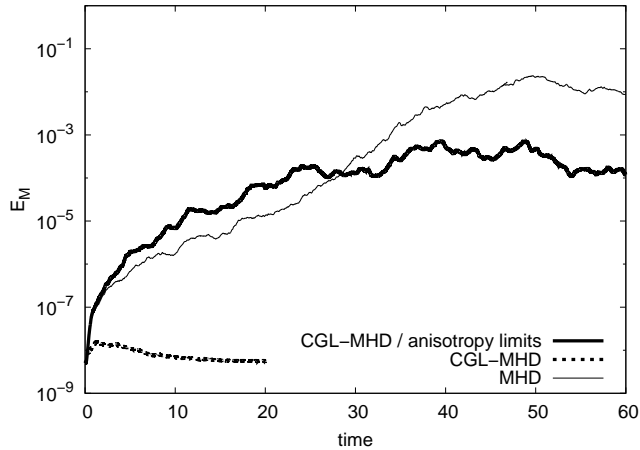} }
\caption[]{Evolution of the magnetic energy for the CGL-MHD turbulent dynamo. Here we compare two models: with and without the anisotropy limits (see equations \ref{eqn:limitanis}). The ordinary MHD case is showed for comparison. Numerical resolution is $64^3$. }
\label{fig1}
\end{figure}

Figure \ref{fig1} shows the evolution of the magnetic energy for two models: one without any limit to the anisotropy, and another imposing the thresholds of equations (\ref{eqn:limitanis}). For comparison, the curve for an ordinary MHD model, employing similar parameters (with an isothermal equation of state) is showed. For our employed parameters, the growth rate of the magnetic field for the CGL-MHD model with limited anisotropy is similar to the MHD turbulent dynamo. However, the magnetic energy of the CGL-MHD model seems to saturate in a value about two orders of magnitude smaller than in the MHD case. For the CGL-MHD model without limited anisotropy, the dynamo fails and the magnetic energy does not grow at all. We also observe failure of the dynamo in simulations employing a KMHD model with isothermal equations of state for the pressure components, whenever that $p_{\perp} > p_{\parallel}$. The suppression of the growth of the magnetic energy by the mirror instability was also observed in \cite[Sharma \etal\ (2006)]{Sharma_etal2006}, in the context of magneto-rotational instability. 

\section{Conclusions and perspectives}

We have showed that the turbulent CGL-MHD dynamo can amplify magnetic energy in a rate similar to the MHD turbulent dynamo --- provided we limit the anisotropy of the pressure --- although the energy of saturation of the magnetic field is smaller. It remains to study the influence of the resolution on the results, as so as the influence of the values of the limits for the anisotropy. Additionally, the structure of the magnetic field in the saturated state for the CGL-MHD model has to be studied. Eventually, the results of the CGL-MHD model have to be compared with observations of the ICM.


\begin{thebibliography}{}

\bibitem[Chew \etal\ (1956)]{Chew_etal1956} 
{Chew, G.~F., Goldberger, M.~L., 
\& Low, F.~E.} 1956, Royal Society of London Proceedings Series A, 236, 112 

\bibitem[En{\ss}lin 
\& Vogt (2006)]{EnsslinVoght2006} 
{En{\ss}lin, T.~A., \& Vogt, C.} 2006, \textit{A\&A}, 453, 447 

\bibitem[Kowal \etal\ (2010)]{Kowal_etal2010} 
{Kowal, G., Falceta-Goncalves, D.~A., \& Lazarian, A.} 2010, arXiv:1012.5125 

\bibitem[Kulsrud (1983)]{Kulsrud1983} 
{Kulsrud, R.~M.} 1983, Basic 
Plasma Physics: Selected Chapters, Handbook of Plasma Physics, Volume 1, 1 

\bibitem[Sharma \etal\ (2006)]{Sharma_etal2006} 
{Sharma, P., Hammett, G.~W., Quataert, E., \& Stone, J.~M.} 2006, \textit{ApJ}, 637, 952 


%
%
%
%
%
%
%
%
%
%
%
%
%
%
%
%
%

\end{thebibliography}
\end{document}